\documentclass[12pt]{iopart}

\usepackage{iopams}  
\usepackage{amssymb}
\usepackage{amsthm}
\usepackage{bm}
\usepackage{xcolor}
\usepackage{mathrsfs}
\usepackage{comment}
\usepackage{graphicx}
\usepackage{siunitx}
\usepackage{lmodern}
\usepackage[normalem]{ulem}

\def\trans{{T}}

\def\P{P}

\def\RR{\mathbb{R}}

\def\neigh{\mathcal{N}}

\def\Gammabf{{\bm \Gamma}}
\def\r{{\bm r}}
\def\W{{\bm W}}

\def\Sigmabf{{\bm \Sigma}}

\def\pot{\mathcal{E}}

\def\Fc{\mathcal{F}}
\def\Fcd{\Fc^{\mathcal{D}}}
\def\Fcr{\Fc^{\mathcal{R}}}

\def\z{\bm z}
\def\p{\bm p}

\def\td{\tilde{r}}
\def\tr{\tilde{\bm r}}
\def\tz{\tilde{z}}
\def\at{\sigma}
\def\atbf{{\bm \at}}
\def\thetabf{{\bm \theta}}
\def\couple{\mathcal{C}}
\def\nobs{N_{{\rm obs}}}
\def\jj{k}
\def\nat{N_{{\rm at}}}
\def\model{\mathcal{M}}

\begin{document}

\title[Equivariant Representation of Friction Tensors]{Equivariant Representation of Configuration-Dependent Friction Tensors in Langevin Heatbaths}

\author{Matthias Sachs$^{\ddagger}$, Wojciech G. Stark$^{\dagger}$, Reinhard J. Maurer$^{\dagger,\$}$, Christoph Ortner$^\#$}
\address{$^{\ddagger}$ School of Mathematics, University of Birmingham, Birmingham B15 2TT, United Kingdom}
\address{$^{\dagger}$Department of Chemistry, University of Warwick, Gibbet Hill Rd, Coventry, CV4 7AL, UK}
\address{$^\$$Department of Physics, University of Warwick, Gibbet Hill Rd, Coventry, CV4 7AL, UK}
\address{$^\#$University of British Columbia, Vancouver, BC, V6T 1Z2, Canada}

\vspace{10pt}

\begin{abstract}
Dynamics of coarse-grained particle systems derived via the Mori-Zwanzig projection formalism commonly take the form of a (generalized) Langevin equation with configuration-dependent friction and diffusion tensors. In this article, we introduce a class of equivariant representations of tensor-valued functions based on the Atomic Cluster Expansion (ACE) framework that allows for efficient learning of such configuration-dependent friction and diffusion tensors from data. Besides satisfying the correct equivariance properties with respect to the Euclidean group E(3), the resulting heat bath models satisfy a fluctuation-dissipation relation. Moreover, our models can be extended to include additional symmetries, such as momentum conservation, to preserve the hydrodynamic properties of the particle system. We demonstrate the capabilities of the model by constructing a model of configuration-dependent tensorial electronic friction calculated from first principles that arises during reactive molecular dynamics at metal surfaces.
\end{abstract}

%
\noindent{\it Keywords}: Langevin equation, electronic friction tensor, density functional theory, dynamics at metal surfaces, atomic cluster expansion
%
%
%
%

\section{Introduction}

Dissipative dynamics governed by a Langevin equation play a crucial role in a wide variety of physical systems such as system-bath models, including dissipative particle dynamics \cite{groot_dissipative_1997} and models of atomistic chemical dynamics in equilibrium with phononic or electronic heat baths. In virtually all realistic systems, the system-bath coupling is such that it gives rise to anisotropic friction with kinematic couplings that depend on the configuration of the system. In addition to the challenge of formulating such general friction tensors, large-scale dynamics simulations require their efficient computation at arbitrary system configurations. 



The Langevin equations arise as the result of a coarse-graining procedure where the number of degrees of freedom of a (not necessarily classical) Hamiltonian system is reduced by a Mori-Zwanzig-like projection with some degrees of freedom of the original system being explicitly modeled by the particle positions and momenta and the remaining degrees of freedom being replaced by the heat bath. An example use of such an approach concerns reactive scattering of a molecule on metal surfaces~\cite{box_determining_2021} where electronic excitations are represented through a configuration-dependent heat bath that gives rise to  {\em electronic friction}. This acts as a first-order correction to the Born-Oppenheimer approximation in molecular dynamics~\cite{Head-Gordon1995}. The resulting friction tensor is positive semi-definite and generally depends on all system coordinates~\cite{Askerka2016}. Formally, this model is a Mori-Zwanzig-type projection of a non-classical electronic Hamiltonian of the atomistic system onto the nuclei's positions and momenta of the adsorbate and metal atoms \cite{dou2017born,dou2018perspective}.

In practice, explicit forms of the potential energy function, the dissipative and the random force of coarse-grained systems are typically not available or are computationally too costly to evaluate for simulation of the Langevin equation over sufficiently long time-scales. In the example case of electronic friction, the corresponding energy function $\pot$ and the electronic friction tensor in the dissipative force are obtained from Density Functional Theory (DFT) calculations for which each evaluation may take several hours to complete~\cite{box_ab_2023}. This computational bottleneck motivates the machine learning of the energy function and the heat-bath in the form of (parametric) surrogate models 
from data that capture the configurational dependence.

Machine learning of accurate and computationally efficient parametric surrogate models, which, in the context of atomistic systems, are commonly referred to as machine-learned interatomic potentials (MLIPs), has been a very active field of research in recent years. Most recent approaches include parametric models based on a linear expansion of the potential functions such as the Atomic Cluster Expansion (ACE)  \cite{drautz2019atomic,dusson2022atomic}, closely related kernel-based methods such as \cite{bartok2010gaussian,bartok2013representing}, as well as 
Graph Neural Network (GNN)-based models such as NequIP \cite{batzner20223}, SchNet \cite{schutt2017schnet}, and MACE \cite{batatia2022mace}.

Reproducing the dynamics of such a coarse-grained system accurately requires not only an accurate surrogate model of the potential energy (MLIP) but also an accurate surrogate model of the heat bath. In contrast to the rich body of literature on representing potential energy functions, there is only limited amount of work on surrogate models of friction tensors. One such model for electronic friction based on a high-dimensional neural network representation was previously proposed and applied to study vibrationally inelastic scattering of nitric oxide at a gold surface~\cite{zhang2019symmetry, box_determining_2021}. This model is based on symmetry-invariant atom-centered features and their derivatives, which are subsequently combined to construct a global matrix-valued output layer that satisfies the relevant algebraic properties of the friction tensor. While the model was able to accurately represent the configurational dependence and symmetry of the friction tensor in this case, a large number of training data points was required. Furthermore, the global output layer construction means that the model is not straightforwardly transferable to larger system sizes and higher dimensional tensors.

In this article, we introduce a novel and general approach 
for representing and learning the heat bath in an underdamped Langevin equation with configuration-dependent anisotropic fluctuation and dissipation. 
The approach that we propose improves upon the above mentioned work in several ways:
\begin{enumerate}
\item the resulting surrogate heat bath model is {\em size-transferable}  in the sense that the model can be evaluated and trained on configurations of arbitrary numbers of particles, 
\item 
the resulting surrogate heat bath model is represented using an equivariant version of the Atomic Cluster Expansion (ACE), which allows for systematic convergence of model accuracy while maintaining computational efficiency, 

\item the model is designed to allow for sparse representation in the case of finite-range interactions, making it scalable to large systems; 
\end{enumerate}

Recently, \cite{lyu2023construction} proposed an approach for learning and representing non-Markovian many-body dissipative particle dynamics (DPD) models. The Markovian analogue of their model has similar features to our approach in that it allows incorporation of many-body effects and is size-transferable. However, their model construction is different from ours in that it is specifically adapted to the structure of dissipative particle dynamics and many-body order interactions are captured through a transformation by learnable non-linear encoder functions. Our approach makes weaker assumptions on the structure of the friction tensor and incorporates many-body order effects on the level of descriptors, which allows for a direct control of the order of the many-body interactions.

The remainder of this article is organized as follows. In section \ref{sec:methods} we introduce the generic structor of our model approach (subsection \ref{sec:generic-model}) and the specific implementation in terms of an equivariant-ACE expansion (subsection \ref{sec:equivariant-expansion}), and model fitting to data (subsection \ref{sec:model-fitting}). Section \ref{sec:results} demonstrates the application of our approach in the example of electronic friction experienced by molecular scattering of nitric oxide (NO) on gold~\cite{box_determining_2021}. We show that the linear model provides prediction accuracy \textit{en par} with the previous neural-network-based approach. The Appendix of this article includes mathematical details on the equivariant-ACE expansion employed in this article and discusses generalizations of the presented noise-coupling approach, and extension and specialization of the proposed heat bath representation to incorporate additional symmetries such as momentum conservation for correct coarse-grained simulation of hydrodynamics.

\section{Methods}\label{sec:methods}

Let $\atbf_{i} = (\r_{i},z_{i},\p_{i})$ denote the state of an atom $i$, which is characterized by its position $\r_{i} \in \mathbb{R}^{3}$, momentum $\p_{i}\in \mathbb{R}^{3}$, and some additional discrete attribute $z_{i}$, in our setting characterizing the atom's chemical element type.
 A tuple $(\atbf_{i})_{i=1}^{\nat}$ of $\nat$ atom states forms a particle system. The following formalism is general for arbitrary particle dynamics. In the application example, we will focus on atoms and molecular dynamics.

The dynamics of atoms in contact with a stochastic heat bath is commonly modelled by a stochastic differential equation of the generic form
\begin{equation}\label{eq:ld}
\eqalign{\dot{\r}_{i} & = {M}_{i}^{-1}\p_{i},  \cr
\dot{\p}_{i} &=  -\nabla_{\r_{i}} \pot((\r_{\jj},z_{\jj})_{\jj}) +  \Fcd_{i}( (\atbf_\jj)_{\jj})+\Fcr_{i}( (\r_{\jj},z_{\jj})_{\jj}).} 
\end{equation}
Here, $\pot$ denotes a potential of mean force, $M_{i}>0$ is the mass of the particle $i$, and the terms $ \Fcd_{i}, \Fcr_{i}$ denote, respectively, the {\em dissipative force} and the {\em random force} experienced by atom $i$. Together these forces model the interaction of the system with a heat bath. Concretely, we assume throughout that (\ref{eq:ld}) takes the form of an underdamped Langevin equation, i.e.,
 \begin{eqnarray}
 \Fcd_{i}( (\atbf)_{\jj}) &=& -\sum_{j=1}^{\nat} \Gammabf_{ij}((\r_{\jj}, z_{\jj})_{\jj})\p_{j}/M_{j},\label{eq:def-df} \\
  \mathcal{F}^{\mathcal{R}}_{i}((\r_{\jj},z_{\jj})_{\jj})  &=& \sqrt{2\beta^{-1}}\sum_{q}\Sigmabf_{iq}((\r_{\jj},\z_{\jj})_{\jj}) \dot{\W}_{iq},\label{eq:def-rf}
 \end{eqnarray}
where $ \dot{\W}_{iq}$ are $m\in \mathbb{N}$ white-noise vectors. The terms $\Gammabf_{ij}((\r_{\jj}, z_{\jj})_{\jj}) \in \mathbb{R}^{3 \times 3}$ and $\Sigmabf_{iq}((\r_{\jj},z_{\jj})_{\jj}) \in \mathbb{R}^{3\times m}$ are atom-wise blocks that make up the {\em friction tensor} $\Gammabf = [\Gammabf_{ij}]_{i,j}$ and {\em diffusion tensor} $\Sigmabf = [\Sigmabf_{iq}]_{i,q}$. At equilibrium, they must satisfy a {\em fluctuation-dissipation} relation of the form
\begin{equation}\label{eq:fd:1}
\Gammabf_{ij} = {\mathbb{E}}[ \mathcal{F}^{\mathcal{R}}_{i} \otimes \mathcal{F}^{\mathcal{R}}_{j}],
\end{equation}
where $\otimes$ denotes the outer product operator and the expectation ${\mathbb{E}}$ is taken with respect to the white noise vectors $\dot{\bm W}_{iq}$. 
Although the methods we describe are fairly general, we will primarily be concerned with the setting where $\Gammabf$ is an electronic friction tensor in equilibrium calculated from first principles theory~\cite{box_ab_2023}.






\subsection{Size-transferable equivariant models for Langevin heat baths}\label{sec:generic-model}

As a consequence of the fluctuation-dissipation relation, the friction tensor must evaluate to a symmetric positive semi-definite matrix for any atomic configuration $(\r_{\jj}, z_{\jj})_{\jj}$. To ensure that this constraint is satisfied, our approach employs an expansion of the diffusion tensor $[\Sigmabf_{iq}]_{i,q}$ and obtains the friction tensor $[\Gammabf_{ij}]_{i,j}$ trough the fluctuation-dissipation relation (\ref{eq:fd:1}), i.e., 
\begin{equation}\label{eq:fd:2}
\Gammabf_{ij} = \sum_{q,q'}\Sigmabf_{iq} \mathbb{E}\left [\dot{\W}_{iq} \dot{\W}_{jq'}^{\trans}\right ]  \left ( \Sigmabf_{jq'}\right )^{\trans}.
\end{equation}
In order for our approach to be size-transferable, we consider the friction tensor to have a functional form where the blocks $\Gammabf_{ij}$ are functions of local coordinates associated with the respective atom pair $(i,j)$, and, in order for such a localization to be physical,  this functional form must reduce to a function of atom states within a local environment of the atom pair $(i,j)$ when imposing finite range interactions between atoms. 
We propose a representation of blocks $\Sigmabf_{iq}$ of the diffusion tensor that results in a friction tensor that is consistent with these requirements by identifying the $q$ index with atoms, i.e., $q \in \{1,\dots,\nat\}$, and imposing a {\em pair coupling} on the white noise vectors $\dot{\W}_{ij}$, i.e., 

\begin{equation}\label{eq:pw-coupling}
 {\mathbb{E}}[\dot{\W}_{iq} \dot{\W}_{jq'}^{\trans}] = \delta_{i q' } \delta_{jq} {\bf I}_{m},
\end{equation}
where $\delta_{ij}$ denotes the Kronecker delta
and ${\bf I}_{m}$ is the $m\times m$ identity matrix. Under this choice of indexing and coupling, the friction tensor takes the concrete form 
\begin{equation}\label{eq:fd:3:pw}
\Gammabf_{ij} = \cases{
\Sigmabf_{ij} \left (\Sigmabf_{ji} \right)^{\trans},& $i \neq j$,\\
\sum_{k=1}^{\nat}\Sigmabf_{ik} \left (\Sigmabf_{ik}\right)^{\trans}, & $i = j$,\\
}
\end{equation}
which indeed can be understood as a function of an atomic pair environment of the atom pair $(i,j)$, provided that each block $\Sigmabf_{kl}$ of the diffusion tensor is a function of a pair environment of the atom pair $(k,l)$. 


In addition to possessing the correct localization properties, blocks $\Sigmabf_{ij}$ of the diffusion tensor must be constructed in such a way that the resulting dissipative forces $\Fcd_{i}$ 
are vector-equivariant with respect to rotations and reflections of inputs, and equivariant under permutation of the particle indexing. These equivariance properties of $\Fcd_{i}$ are equivalent (see  \ref{ap:sec:derivation:gamma}) to the following symmetries of blocks $\Gammabf_{ij}$ of the friction tensor:
 \begin{eqnarray}
\forall Q \in O(3): \quad  &\Gammabf_{ij}((Q\r_{\jj}, z_{\jj})_{\jj}) &= Q\Gammabf_{ij}((\r_{\jj}, z_{\jj})_{\jj})Q^{\trans},\label{eq:gamma:equi}\\
\forall \sigma \in S_{\nat}: \quad  &\Gammabf_{ij}(\sigma \cdot (\r_{\jj}, z_{\jj})_{\jj}) &= \Gammabf_{\sigma i,\sigma j}((\r_{\jj}, z_{\jj})_{\jj}).\label{eq:gamma:permeq}
\end{eqnarray}

To ensure permutation equivariance in the sense of (\ref{eq:gamma:permeq}), the diffusion tensor blocks $\Sigmabf_{ij}$ must be invariant under all permutations $\sigma \in S_{\nat}$ that leave the indices of the atom pair $(i,j)$ unchanged, i.e.,
\begin{equation}\label{eq:restricted-perm-inv}
\forall \sigma \in S_{\nat}, \sigma j = j, \sigma i = i : \quad \Sigmabf_{ij}(\sigma \cdot (\r_{\jj}, z_{\jj})_{\jj}) = \Sigmabf_{ij}((\r_{\jj}, z_{\jj})_{\jj}).
\end{equation}
In combination with the pair-wise coupling (\ref{eq:pw-coupling}) of the white-noise vectors, permutation equivariance (\ref{eq:gamma:permeq}) is ensured as detailed in \ref{app:coupling-condition}. Other couplings that ensure permutation equivariance can be constructed and we refer for a discussion of such alternative couplings to \ref{app:alternative-coupling}.

To ensure that $O(3)$-equivariance (\ref{eq:gamma:equi}) is satisfied we let the blocks of the diffusion tensor be comprised of a sclar-equivariant sub-block $\Sigmabf^{(0)}_{ij}$ (transforms like a scalar, i.e., entries are invariant under rotations and reflections), vector-equivariant sub-block $\Sigmabf^{(1)}_{ij}$ (transforms like a covariant vector) and a matrix-equivariant sub-block $\Sigmabf^{(2)}_{ij}$, (transforms like a matrix). That is, we let 
\begin{equation}\label{eq:dt-decomp}
\Sigmabf_{ij} = \left [\Sigmabf^{(0)}_{ij}, \Sigmabf^{(1)}_{ij}, \Sigmabf^{(2)}_{ij} \right],
\end{equation}
where $\Sigmabf^{(l)}_{ij} =  [\Sigmabf^{(l,p)}_{ij}]_{p=1}^{m_{l}}$, with $m= 3m_{0} + m_{1}+3m_{2}$ and the components 
\[
\hspace{-1cm}
\Sigmabf^{(0,p)}_{ij}((\r_{\jj}, z_{\jj})_{\jj}) \in \mathbb{R}^{3\times 3}, \quad  \Sigmabf^{(1,p)}_{ij}((\r_{\jj}, z_{\jj})_{\jj}) \in \mathbb{R}^{3}, \quad  \Sigmabf^{(2,p)}_{ij}((\r_{\jj}, z_{\jj})_{\jj}) \in \mathbb{R}^{3\times 3},
\]
satisfy the symmetries
\begin{eqnarray}
 \forall Q \in O(3): \quad  &\Sigmabf^{(0,p)}_{ij}((Q\r_{\jj}, z_{\jj})_{\jj})&=\Sigmabf^{(0,p)}_{ij}((\r_{\jj}, z_{\jj})_{\jj}),\\
\forall Q \in O(3): \quad  &\Sigmabf^{(1,p)}_{ij}((Q\r_{\jj}, z_{\jj})_{\jj})&=Q\Sigmabf^{(1,p)}_{ij}((\r_{\jj}, z_{\jj})_{\jj}),\\
\forall Q \in O(3): \quad  &\Sigmabf^{(2,p)}_{ij}((Q\r_{\jj}, z_{\jj})_{\jj})&=Q\Sigmabf^{(2,p)}_{ij}((\r_{\jj}, z_{\jj})_{\jj})Q^{\trans}.
\end{eqnarray}
We note that equivariance of the friction tensor (\ref{eq:gamma:equi}) implies that $\Sigma_{ij}^{(0,p)} = \zeta_{ij}((\r_{\jj}, z_{\jj})_{\jj}) {\bf I}_3$, where $\zeta_{ij}$ is an $O(3)$-invariant scalar.

Excluding all matrix-equivariant components $\Sigmabf^{(2,p)}_{ij}$ would result in an incomplete class of friction models. For example, a friction tensor of the form $\Gammabf_{ij}((\r_{\jj}, z_{\jj})_{\jj}) = \gamma((\r_{\jj}, z_{\jj})_{\jj}) {\bf I}_{3}$, where $\gamma$ is invariant scalar-valued, cannot be represented using a continuous diffusion tensor $[\Sigmabf_{ij}]_{i,j}$ that is comprised of only vector-equivariant components (cf. ``hair ball theorem''). Vector-equivariant components, $\Sigmabf^{(1,p)}_{ij}$ can in theory always be replaced with a matrix-equivariant component (see \ref{sec:app:mv-equ}), but we include those in our model approach since the additional computational cost is negligible and they allow for more efficient representation and expansion of certain low-rank friction tensors. Such low-rank friction tensors are for example relevant in dissipative particle dynamics where friction and diffusion matrix need to satisfy momentum-conservation; see, e.g., \cite{espanol2017perspective}.

Similarly, since invariant components $\Sigmabf^{(0,p)}_{ij}$ must be of the functional form $\Sigmabf^{(0,p)}_{ij}((\r_{\jj}, z_{\jj})_{\jj}) = \zeta_{ij}((\r_{\jj}, z_{\jj})_{\jj}) {\bf I}_{3}$, it follows that every invariant component is also matrix-equivariant and can thus be replaced by a general matrix-equivariant component $\Sigmabf^{(2,p)}_{ij}$. Yet, in applications where the friction tensor is a-priori known to be of the structural form $[\gamma_{ij}^{2}((\r_{\jj}, z_{\jj})_{\jj}) {\bm I}_{3}]_{i,j}$ where $\gamma_{ij}$ is invariant, scalar-valued, it would be prudent to include only scalar-invariant components and no general matrix-equivariant components in the friction model as the latter result in unnecessary model complexity.

\subsection{Local equivariant expansion of diffusion tensor blocks}\label{sec:equivariant-expansion}
%
%
To account for the restricted permutation invariance condition (\ref{eq:restricted-perm-inv}), we perform local expansions of blocks of the diffusion tensor on environments of atom pairs rather than on atomic site environments as commonly employed in the construction of MLIPs. Here we present the general structural form of such an expansion and its implementation in the linear framework of the ACE; details are deferred to \ref{sec:pair-env-expansion}. 

\begin{figure}
\begin{center}
\includegraphics[width=.9\linewidth]{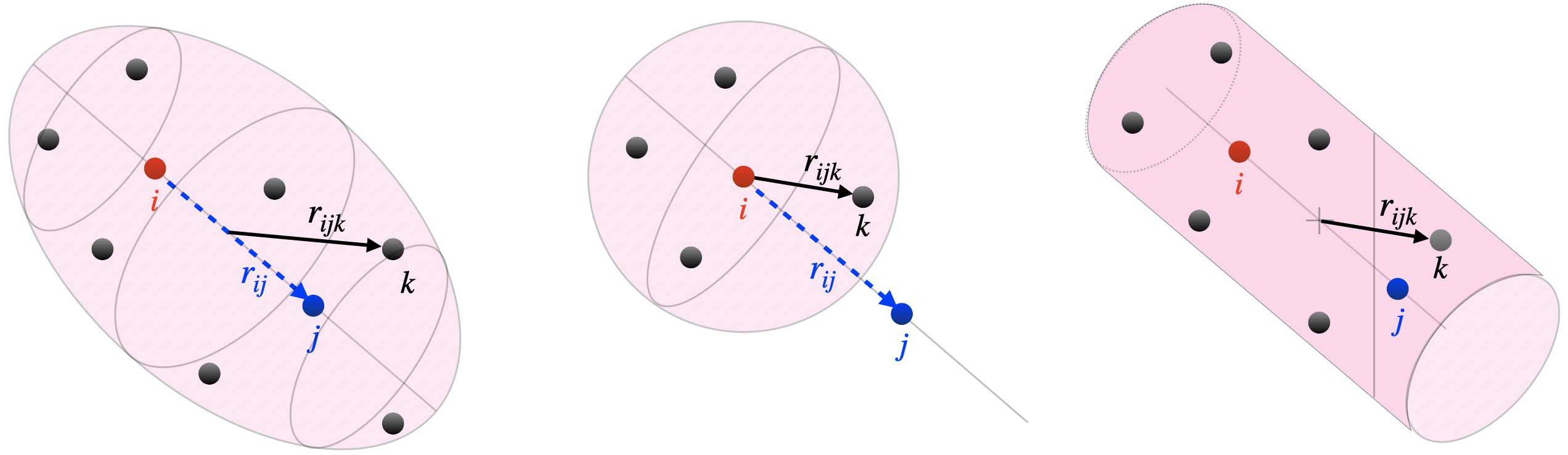}
\caption{Illustration of different pair-environments. From left to right: Ellipsoid cutoff $E_{\lambda, \mu, r_{\rm cut}}$ with $\lambda=1/2$ and $\mu>1$; 
Orientated spherical cutoff corresponding to an ellipsoid cutoff $E_{\lambda, \mu, r_{\rm cut} }$ with $\lambda=0$ and $\mu=1$; Cylindrical cutoff as employed in~\cite{zhang2022equivariant}. } \label{fig:env-examples}
\end{center}
\end{figure}

We refer to $\neigh(i,j)
:= \{ k ~: \r_{k} \in E(\r_{i},\r_{j})\}$ as the neighbourhood of the atom pair $(i,j)$, where $E(\r_{i},\r_{j})$, termed the {\em cutoff function}, defines a sub-domain of $\RR^{3}$ that is invariant under rotations about the axis $\r_{ij} := \r_{j}-\r_{i}$. Examples of such atom pair neighbourhoods are shown in Figure~\ref{fig:env-examples}. 
In particular, in this work, we consider cutoff domains of ellipsoid shape, centered at a position $ \r_{ij}(\lambda) := (1-\lambda) \r_{i} + \lambda \r_{j}, \lambda \in [0,1]$ on the line connecting the atoms $i$ and $j$ and major-axis aligned with the displacement orientation  $ \hat{\r}_{ij} := \r_{ij}/ |\r_{ij}|_{2}$ of the atoms $i$ and $j$, i.e.,
\begin{equation}
E_{\lambda, \mu, r_{\rm cut} }(\r_{i},\r_{j}) :=\left\{ \r \in \RR^{3}:    \frac{\left | \mathcal{P}  \r  \right |^{2}}{\mu^{2}} +  \left | \r - \mathcal{P} \r  \right |^{2} < r^{2}_{\rm cut} \right \},
\end{equation}
where 
\begin{equation}\label{eq:proj-center}
\mathcal{P} \r := (\r - \r_{ij}(\lambda) ) \cdot \hat{\r}_{ij}, 
\end{equation}
denotes the projection onto the major axis of the ellipsoid, the parameter {$r_{\rm cut} > 0$} corresponds to the maximum cutoff distance along the minor axes and $\mu\geq 1$ determines the maximum cutoff distance, $\mu r_{\rm cut}$, along the major axis. 

\begin{figure}
\begin{center}
\includegraphics[width=.35\linewidth]{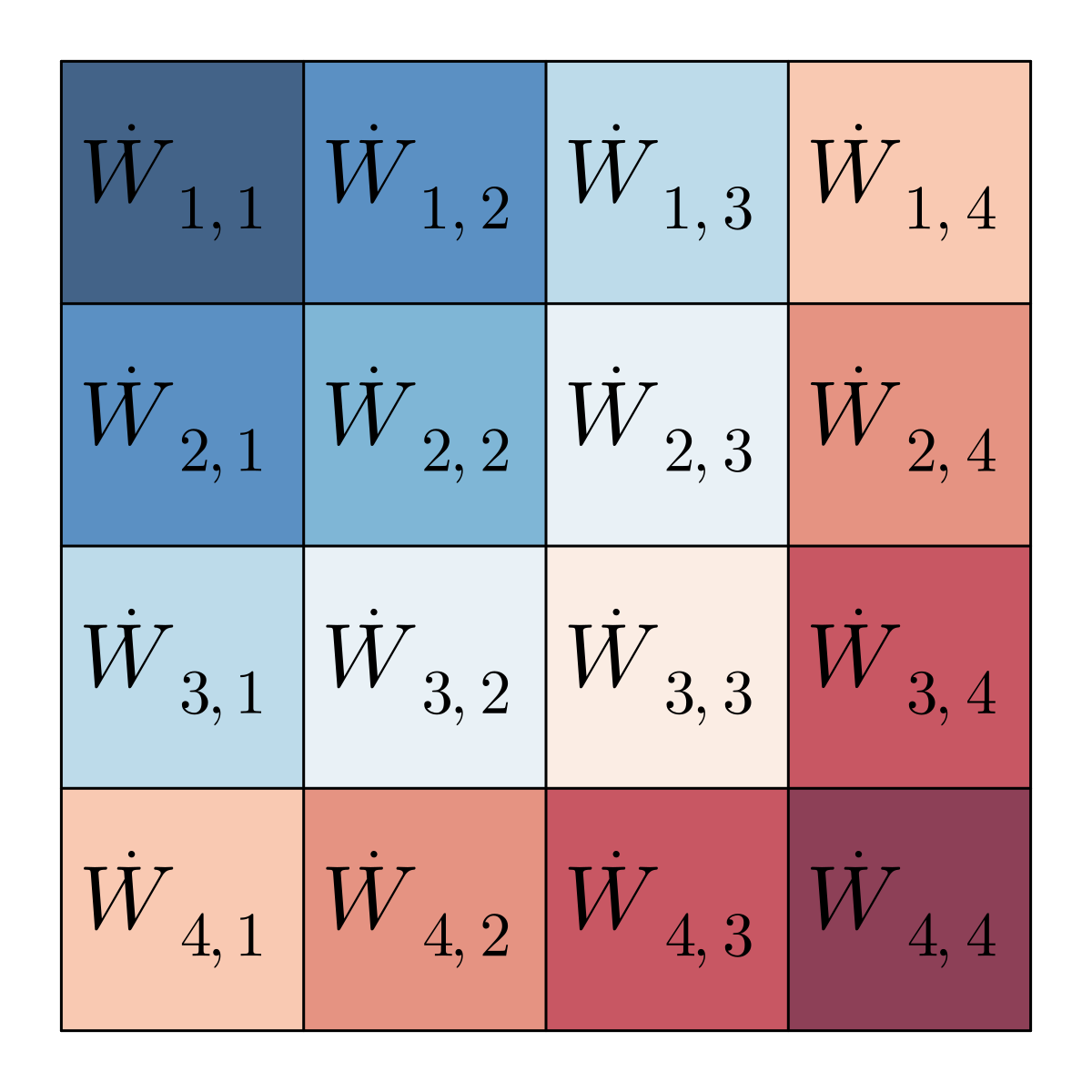}
\caption{Illustrations of the pair-wise coupling in the case of a 4-particle system. Noise vectors with same background color are identical and thus coupled. Noise vectors with different background colors are independent. 
} 
\end{center}
\end{figure}
To allow representation of blocks of the diffusion tensor by functions of local coordinates, we restrict the functional forms for $\Sigmabf^{(l,p)}_{ij}$ to 
\begin{equation}\label{eq:model-func-form}
\hspace{-2cm} \Sigmabf^{(l,p)}_{ij}\left((\r_{k},\z_{k})_{k}\right) = \cases{ \model^{(l,p)} \left (z_i, z_j,\r_{ij}, \{ (\r_{ijk}, z_{k})\}_{k \in \neigh(i,j)}\right ), & $| \r_{ij} | < e_{\rm cut}$,\\
{\bf 0 }, & \text{otherwise}.}
\end{equation}
where $\r_{ijk} = \r_k -  \r_{ij}(\lambda)$, and some prescribed $e_{\rm cut}>0$. The model functions
$\model^{(l,p)}$ satisfy the same symmetries as $\Sigmabf^{(l, p)}_{ij}$, which translate to 
\begin{eqnarray}
\label{eq:symM0p}
\hspace{-2.3cm} \model^{(0,p)} \left (z_{i},z_{j}, Q\r_{ij}, \{ Q\r_{ijk}, z_{k}\}_{k\in \neigh(i,j)}\right ) 
&= \model^{(0,p)} \left (z_{i}, z_{j}, \r_{ij}, \{ \r_{ijk}, z_{k}\}_{k\in \neigh(i,j)}\right ),  \\
\label{eq:symM1p}
\hspace{-2.3cm} \model^{(1,p)} \left (z_{i},z_{j}, Q\r_{ij}, \{ Q\r_{ijk}, z_{k}\}_{k\in \neigh(i,j)}\right ) 
&= Q\model^{(1,p)} \left (z_{i}, z_{j}, \r_{ij}, \{ \r_{ijk}, z_{k}\}_{k\in \neigh(i,j)}\right ),  \\
\label{eq:symM2p}
\hspace{-2.3cm} \model^{(2,p)} \left (z_{i}, z_{j},Q\r_{ij}, \{ Q\r_{ijk}, z_{k}\}_{k\in \neigh(i,j)}\right ) 
&= Q\model^{(1,p)} \left (z_{i},z_{j}, \r_{ij}, \{ \r_{ijk}, z_{k}\}_{k\in \neigh(i,j)}\right )Q^{T}.
\end{eqnarray}
The permutation equivariance condition (\ref{eq:restricted-perm-inv}) is guaranteed by invariance of $\model^{(l,p)}$ under relabelling of atoms in the pair neighborhood  $\neigh(i,j)$, which is implicit in treating the environment as a multi-set, and by treating the displacement $\r_{ij}$ between the atom $i$ and atom $j$ as a separate argument, which allows to restrict permutation invariance to permutations that leave the indices $i,j$ unchanged.

There is now a wide range of model architectures available that can be modified or extended to parameterize $\model^{(l, p)}$; see e.g., \cite{batatia2022mace,villar2021scalars,batzner20223,geiger2022e3nn} for architectures that use internal equivariant representations, and \cite{drautz2020atomic,zhang2022equivariant,nigam2022equivariant,grisafi2018symmetry} for architectures with equivariant outputs, as well as references therein. 
We employ the equivariant atomic cluster expansion \cite{drautz2020atomic}, extended to pair environments in \cite{zhang2022equivariant}, adapted to our setting. This results in a linear model for the diffusion tensor blocks $\Sigmabf^{(l,p)}_{ij}$, 
\begin{equation}
    \label{eq:param_Mlp}
    \model^{(l, p)}(\,\bullet\,) = {\bm c}^{(l, p)} \cdot {\bm B}^{(l)}(\,\bullet\,)
\end{equation}
where ${\bm B}^{(l)}$ is a vector of basis functions (or, dictionary) that satisfy the symmetries (\ref{eq:symM1p}, \ref{eq:symM2p}) of $\model^{(l,p)}$  exactly and ${\bm c}^{(l, p)}$ the corresponding vector of (scalar) parameters. 
The main model hyperparameters that will be explored in the next section are the cutoff radius for the interaction, the maximum correlation order (body-order minus~1), and the maximum total polynomial degree of the features from which the basis is constructed.
The details of the basis construction and of the aforementioned hyperparameters are presented in~\ref{app:pair-ace}.

\subsection{Model fitting}\label{sec:model-fitting}
We collect all parameters into a single parameter vector ${\bm \theta} = ({\bm c}^{(l, p)})_{l, p}$. We remark that the model for the diffusion tensor $\Sigmabf_{ij}$ resulting from (\ref{eq:param_Mlp}) is linear in ${\bm \theta}$, but the resulting parameterized model for the friction tensor $\Gammabf_{ij}^{\bm \theta}$ is quadratic in ${\bm \theta}$.

We estimate the parameters ${\bm \theta}$ to a dataset comprised of {$\nobs$} atomic structures $R^{(n)} =  (\r_{i}^{(n)}, z_{i}^{(n)})_{i=1}^{\nat}$ and target friction tensors $\Gamma^{(n)} = (\Gamma^{(n)}_{ij})_{i,j=1}^{\nat}$ obtained from a reference model (see below for details). This is achieved by minimizing the least squares loss
\begin{equation}
\mathcal{L}({\bm \theta}) = \sum_{n=1}^{{\nobs}} \sum_{i,j} \left | \Gamma^{(n)}_{ij} - \Gammabf^{\thetabf}_{ij}(R^{(n)}) \right|_{2}^{2}, 
\end{equation}
where $|\cdot|_2$ denotes the Frobenius matrix norm. Since $\Gammabf^{\thetabf}$ is quadratic in $\thetabf$, the loss function $\mathcal{L}(\thetabf)$ is quartic in $\thetabf$ and must therefore be optimized using an iterative nonlinear optimizer. To obtain the model presented in this work, we used the ADAM optimizer \cite{kingma2014adam} and the Polyak Momentum method \cite{polyak1964some} in combination with batch-sampling of the observation data to minimize the loss function $\mathcal{L}(\thetabf)$. As our model fit indicates (see Figures \ref{fig:optimization_arch_cutoff_pw}-\ref{fig:learning_curve_pw}), the implicit regularization due to the employed batch sub-sampling was sufficient to obtain model fits with robust generalization properties and we thus did not employ explicit additional regularization in our model fits.

\section{Results}\label{sec:results}

\subsection{Data set and model training}
The data set containing electronic friction tensors (EFTs) for NO/Au(111) from~\cite{box_determining_2021} is used in this study. The structures in the data set are based on ab-initio simulations from~\cite{yin_strong_2019} calculated with the Vienna Ab initio Simulation Package (VASP)~\cite{kresse_efficiency_1996,kresse_efficient_1996} using PW91 functional~\cite{perdew_atoms_1992} and a Gamma-centered k-point grid of 4$\times$4$\times$1. EFTs were calculated for the structures in the data set using 4-layered 3$\times$3 metal slabs with  70~$\mathrm{\mathring{A}}$ of vacuum spacing in the $z$-direction. For EFT evaluation, the PBE functional~\cite{perdew_generalized_1996} with a ``tight'' basis set, and a 9$\times$9$\times$1 k-point grid were employed within the FHI-aims code~\cite{blum_ab_2009}. Additionally, the Gaussian broadening function with a width of 0.6~eV was used and the electronic temperature was set at 300~K.
The final training and test sets contained 1647, and 1000 structures, respectively. More details about the data generation and EFT evaluation can be found in~\cite{yin_strong_2019, box_determining_2021}.

The EFT models in this study were trained using Adam optimizer with a learning rate of 10$^{-4}$ and a momentum decay rate $\beta_{1}=0.99$ and decay rate $\beta_{2}=0.999$ for squared gradient averages. 
The parameters of the models are discussed in more detail in the following section.


To assess the accuracy of our fits we evaluate the root mean square error (RSME) 
as the square root of average squared residual error per scalar entry of the friction tensor, i.e.,
\begin{equation}\label{eq:rmse}
    \mathrm{RMSE} := \sqrt{ \frac{1}{\nobs} \sum_{n=1}^{\nobs} \frac{1}{ N_{\rm ent}^{(n)} }  \sum_{i,j} \left | \Gamma^{(n)}_{ij} - \Gammabf^{\thetabf}_{ij}(R^{(n)}) \right|_{2}^{2}}, 
\end{equation}
where $N_{\rm ent}^{(n)}: =  \left (3\nat^{(n)}\right )^{2}$ is the number of scalar entries of the $n$th friction tensor in the dataset.
Similarly, we compute the mean absolute error (MAE) of our fits as the average absolute residual error per scalar entry, i.e., 
\begin{equation}\label{eq:mae}
    \mathrm{MAE} := \frac{1}{\nobs} \sum_{n=1}^{\nobs}\frac{1}{N_{\rm ent}^{(n)} }  \sum_{i,j} \left | \Gamma^{(n)}_{ij} - \Gammabf^{\thetabf}_{ij}(R^{(n)}) \right|_{1},
\end{equation}
where $|{\rm A}|_{1} := \sum_{k=1}^{m} \sum_{l=1}^{m}|a_{kl}|$ for ${\rm A} = [a_{kl}]_{k,l} \in \RR^{m\times m}$.

\subsection{Model optimization}
\begin{figure}
\centering
\includegraphics[width=.9\linewidth]{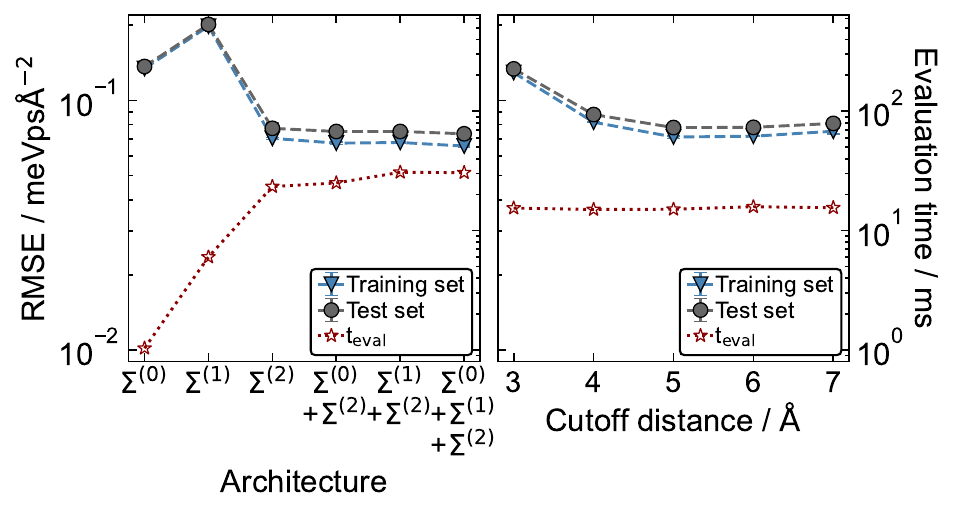}
\caption{\textbf{Optimization of architecture type and cutoff distance with pair-wise coupled models.} The test (dark grey circles), and training (blue triangles) RMSEs obtained using different combinations of invariant ($\Sigmabf^{(0)}$), vector-equivariant ($\Sigmabf^{(1)}$), and matrix-equivariant ($\Sigmabf^{(2)}$) representations of $\Sigmabf$ with cutoff distance of 5.0~$\mathrm{\mathring{A}}$ (left side), and the convergence of EFT with respect to cutoff distance in $\mathrm{\mathring{A}}$ units, employing $\Sigmabf^{(0)}$ and $\Sigmabf^{(2)}$ (right side). Additionally, corresponding average evaluation times are included, based on 100 EFT evaluations (red stars). The maximum correlation order of 2 and corresponding polynomial degree of 5 was used in every model shown in both figures. } 
\label{fig:optimization_arch_cutoff_pw}
\end{figure}

\begin{figure}
\centering
\includegraphics[width=.9\linewidth]{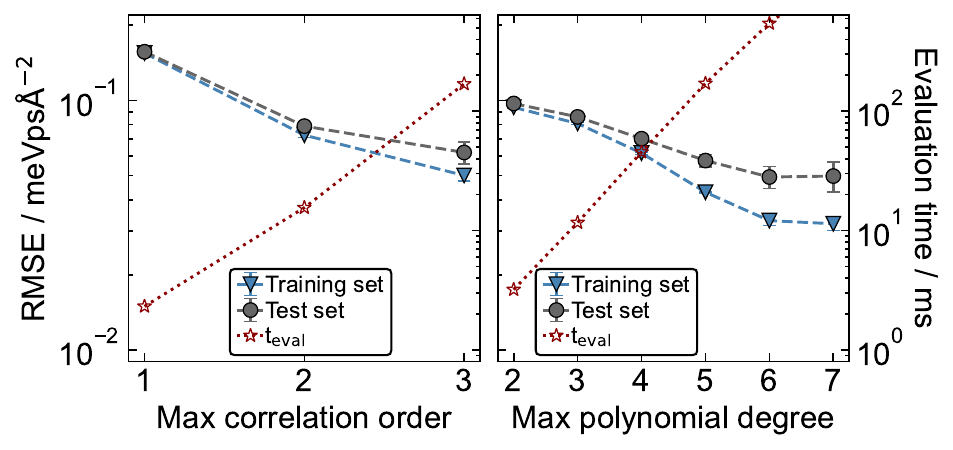}
\caption{\textbf{Effect of maximum correlation order and maximum polynomial degree with pair-wise coupled models.} The test (dark grey circles), and training (blue triangles) RMSEs were obtained for different correlation orders with constant polynomial degree of 5 (left side), and polynomial degrees for a correlation order of 3 (right side). Additionally, corresponding average evaluation times are included, based on 100 EFT evaluations (red stars). Cutoff distance in both plots was set at 6~$\mathrm{\mathring{A}}$, and  invariant ($\Sigmabf^{(0)}$) and matrix-equivariant ($\Sigmabf^{(2)}$) representations were used in training. } 
\label{fig:optimization_ord_degr_pw}
\end{figure}

The optimization of model parameters, based on cross-validation, was performed for the EFT models trained on NO/Au(111) data. In our approach, we split our training data set into 5 equal parts, and we created 5 different splits containing 4 out of 5 parts, in such a way that in each split, a different part of the data is missing. The 5 splits were then used to train 5 separate models for every setting, and the corresponding errors, with the standard deviations of the error obtained by the 5 models, together with the evaluation times obtained by the models, were plotted (Figs.~\ref{fig:optimization_arch_cutoff_pw}~and~\ref{fig:optimization_ord_degr_pw}). Evaluation times were calculated by taking an average of 100 evaluations of an entire EFT.

The impact on prediction accuracy and evaluation time of different combinations of the components $\Sigmabf^{(l)}, l=0,1,2$ in the representation of   $\Sigmabf$, and cutoff distances, is shown in Fig.~\ref{fig:optimization_arch_cutoff_pw}.
The RMSEs of models containing different combinations of invariant, vector-equivariant, and matrix-equivariant components show that matrix-equivariance is the most important property, and enables low RMSEs of EFT prediction (roughly 0.07~meVps$\mathrm{\mathring{A}}^{-2}$), as compared to models that use only invariant or vector-equivariant components (0.1--0.2~meVps$\mathrm{\mathring{A}}^{-2}$). However, including matrix-equivariant components leads to higher model evaluation times (20~ms versus 5~ms with invariant).
The high approximation error (RMSEs of roughly 0.2~meVps$\mathrm{\mathring{A}}^{-2}$) of the model comprised only of vector-equivariant components suggests that those components contribute the least to the representation of the EFT.

Friction models were trained using the pair-wise coupling (\ref{eq:pw-coupling}) and different cutoff distances, between 3 and 7~$\mathrm{\mathring{A}}$. The observed RMSEs improve significantly when the distance is increased from 3 to 4~$\mathrm{\mathring{A}}$, however only slight improvement is visible when increasing the distance further to 5~$\mathrm{\mathring{A}}$. For cutoff distances larger than 5~$\mathrm{\mathring{A}}$, the errors increase slightly, which may be caused by the overlapping periodic images of the cell, or by the need of the ML model to resolve a larger interaction environment.

The effect of the maximum correlation order and polynomial degree of the ACE basis on the model performance is shown in Fig.~\ref{fig:optimization_ord_degr_pw}. EFTs are well approximated when using ACE bases of correlation order 3. The model evaluation time increases roughly 5-fold with every order, thus it may be important to use low, but converged values of the correlation order. The effect of the corresponding maximum polynomial degrees (for correlation order of 3) is shown on the right side of Fig.~\ref{fig:optimization_ord_degr_pw}. The improvement of the model is visible until polynomial degree of 6--7 which provide test RMSEs of roughly 0.033~meVps$\mathrm{\mathring{A}}^{-2}$. The difference between the training and test set RMSEs is more pronounced for larger degrees. The difference between the degrees of 6 and 7 in test RMSEs is negligible. Evaluation times appear to increase exponentially with the increasing polynomial degrees, but this is a pre-asymptotic effect; the actual increase is algebraic (depending on correlation order).


For the final, optimized model, we use a cutoff distance of 5~$\mathrm{\mathring{A}}$, a maximum correlation order of 3, and a corresponding maximum polynomial degree of 7.

\subsection{Friction model performance}
\begin{figure}
\centering
\includegraphics[width=.5\linewidth]{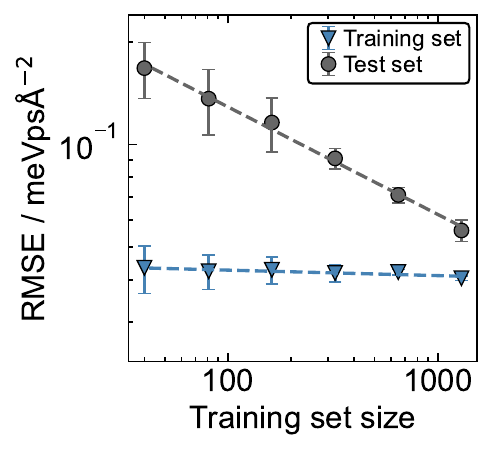}
\caption{\textbf{Learning curve for the pair-wise coupled models.} 
Log-log plot of the averaged test (dark grey circles), and training (blue triangles) set predictive error (RMSE) in the evaluation of friction tensor elements as a function of training set size for NO/Au(111) model. The shown RMSE values are averaged over 5 cross-validation splits, differing by training sets. Error bars correspond to standard deviations between the RMSEs obtained over all splits. All the models were trained using the cutoff distance of 5~$\mathrm{\mathring{A}}$, correlation order of 3 with polynomial degree of 5, and with all 
the invariant ($\Sigmabf^{(0)}$), vector-equivariant ($\Sigmabf^{(1)}$), and matrix-equivariant ($\Sigmabf^{(2)}$) components.} 
\label{fig:learning_curve_pw}
\end{figure}

To illustrate the learning capabilities of our model, we plot the learning curve (Fig.~\ref{fig:learning_curve_pw}), in which prediction errors (RMSEs) are shown for models with varying training set sizes. The test set RMSE for the model trained on 40 structures is slightly above 0.175~meVps$\mathrm{\mathring{A}}^{-2}$, and rapidly improves with more data, approaching the limit of training set RMSEs when the training set includes more than 1000 structures (roughly 0.056~meVps$\mathrm{\mathring{A}}^{-2}$). The training set RMSEs converge instantly and do not improve with the increasing training set size, staying at roughly 0.04~meVps$\mathrm{\mathring{A}}^{-2}$. Further improvement is most likely blocked by the intrinsic numerical error of EFT evaluation within the electronic structure code \cite{box_ab_2023}. This suggests that the optimized models can achieve approximation errors that are comparable in magnitude to errors intrinsic to the reference electronic structure method.

\begin{figure}
\centering
\includegraphics[width=0.9\linewidth]{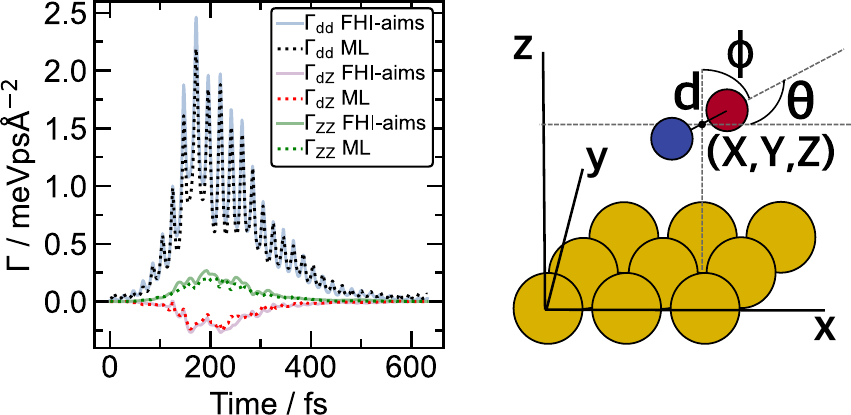}
\caption{\textbf{Predictions of electronic friction tensor elements in internal coordinates ($\Gamma_{\mathrm{dd}}$, and $\Gamma_{\mathrm{dZ}}$, $\Gamma_{\mathrm{ZZ}}$) for NO scattering trajectory at Au(111) surface.} Electronic friction elements $\Gamma_{\mathrm{dd}}$ and $\Gamma_{\mathrm{dZ}}$ (internal coordinates) are plotted for scattering trajectory, as predicted by our model (dotted lines) and compared to reference values (solid lines) obtained with FHI-aims code (left).} 
\label{fig:validation_noau_model}
\end{figure}

To validate the model performance, we evaluated the friction model along a molecular dynamics trajectory, in which a NO molecule initialised in the third excited vibrational state ($\nu=3$) scatters at the Au(111) surface. Fig.~\ref{fig:validation_noau_model} compares three friction tensor elements in internal coordinates ($\Gamma_{\mathrm{dd}}$ and $\Gamma_{\mathrm{dZ}}$) with the FHI-aims reference calculations. Coordinate $dd$ corresponds to the interatomic stretch motion of the atoms in the NO molecule, and $ZZ$ corresponds to the centre-of-mass motion of the molecule with respect to the surface. The coordinate transformation is defined in appendix \ref{sec:app:eft_transform}.  The off-diagonal element, $\Gamma_{\mathrm{dZ}}$, describes the kinematic frictional coupling of the two degrees of freedom. The agreement between the model prediction and the reference results is excellent along the whole trajectory. The model correctly resolves the rapid oscillations in friction due to vibrational motion and the overall magnitude of friction for all three components as the molecule reaches the turning point closest to the surface (at around 200~fs).

The final model test RMSE is 0.046~meVps$\mathrm{\mathring{A}}^{-2}$, and MAE is 0.025~meVps$\mathrm{\mathring{A}}^{-2}$. The reference model from Ref.~\cite{box_determining_2021} using the same data set achieved a very similar test RMSE of 0.045~meVps$\mathrm{\mathring{A}}^{-2}$ (MAE was not reported). Considering, the learning curve in Fig.~\ref{fig:learning_curve_pw}, we can see that further improvement of the models is likely not possible based on the data provided. Therefore, both models reached the lowest possible levels of errors.

\section{Conclusions}
We presented a framework for the construction of equivariant representations of the friction tensor and diffusion tensor in a Langevin heat bath that by construction satisfy necessary symmetries including a fluctuation-dissipation relation. The framework is agnostic to the presented implementation based on the Atomic Cluster Expansion (ACE). Here we use ACE as it provides computationally efficient basis evaluation but other choices are available.

We showcased on the example of NO scattering at Au(111) surfaces that the proposed representation, when implemented using ACE, provides models of greatly reduced complexity and increased computational efficiency at accuracies comparable with the previous most successful NN-based methods. Furthermore, the independent treatment of atom-wise blocks to construct the friction tensor makes the model size-transferable to larger simulations. In the case of electronic friction, that means simulations of multiple adsorbates and larger friction tensors.

The framework can be extended to non-Markovian scenarios in the form of Quasi-Markovian generalized Langevin Equation~\cite{leimkuhler2019ergodic,lim2020homogenization,vroylandt2022position} and coarse-grained models where additional symmetries may apply, for example dissipative particle dynamics \cite{espanol2017perspective,li2017computing,osti_1565466} where momentum-conservation between system and bath must be satisfied. The current model only considers scenarios where the second fluctuation dissipation theorem holds, but nonequilibrium and open systems can also be considered in the future by building independent symmetrized constructions of the friction and diffusion tensors.

\appendix

\section{Derivation of equivariance properties of $\Gammabf$ }\label{ap:sec:derivation:gamma}
The dissipative forces $\Fcd_{i}(\cdot), (i=1,\dots,\nat)$ are vector-equivariant with respect to rotations and point reflections, and equivariant with respect to permutations in the following sense
 \begin{eqnarray}
\forall Q \in O(3): &\Fcd_{i}( Q \cdot \atbf) &=  Q\Fcd_{i}(\atbf),\label{eq:df:roteq}\\
\forall \sigma \in S_{\nat}: &\Fcd_{i}( \sigma \cdot \atbf) &= \Fcd_{\sigma i}(\atbf),\label{eq:df:permeq}
\end{eqnarray}
for any system $\atbf = (\at_{i})_{i}^{\nat}$ of $\nat$ particles. 
 Under the assumption that the dissipative force takes the form (\ref{eq:def-df}), the equivariance relations of (\ref{eq:df:roteq})-(\ref{eq:df:permeq}) can be rewritten as \begin{eqnarray}
\forall Q \in O(3): \quad  &\sum_{j=1}^{\nat} \Gammabf_{ij}((Q\r_{i}, z_{i})_{i}) Q\p_{j} = Q\sum_{j=1}^{\nat}\Gammabf_{ij}((\r_{\jj}, z_{\jj})_{\jj}) \p_{j},\\
\forall \sigma \in S_{\nat}: \quad  &\sum_{j=1}^{\nat} \Gammabf_{ij}((\r_{\sigma \jj}, z_{\sigma \jj})_{\jj}) \p_{\sigma j} = \sum_{j=1}^{\nat}\Gammabf_{ \sigma i \sigma j}((\r_{\jj}, z_{\jj})_{\jj}) \p_{\sigma j},
\end{eqnarray}
respectively, which hold for every system state $(\r_{\jj}, z_{\jj}, \p_{\jj})_{\jj}$ exactly if the equivariance relations  (\ref{eq:gamma:equi}) and (\ref{eq:gamma:permeq}) hold for every configuration $(\r_{\jj}, z_{\jj})_{\jj}$ and all index pairs $k,l$.

\section{Sufficient conditions for equivariance properties of $\Sigmabf$ }\label{ap:sec:derivation:sigma}
We define the {\em coupling function} $\couple: \{1,\dots,\nat\}^{4} \rightarrow \{0,1\}$, of the white-noise vectors $\dot{\W}_{iq}$ through the following relation
\begin{equation}
 {\mathbb{E}}[\dot{\W}_{iq} \dot{\W}_{jq'}^{\trans}] = \couple(i,q,j,q') {\bf I}_{m}.
\end{equation}
Any such defined coupling function satisfies $|\couple(i,q,j,q')| \geq \delta_{ij} \delta_{qq'}$, and the minimal coupling $\couple(i,q,j,q') =\delta_{ij} \delta_{qq'}$ holds if all white-noise vectors are independent.
\subsection{ $O(3)$-Equivariance}
By virtue of the fluctuation-dissipation relation (\ref{eq:fd:2}), the equivariance relation (\ref{eq:gamma:equi}) can be written in equivalent form in terms of the block entries of the diffusion tensor as
\begin{equation}
\eqalign{\forall Q \in O(3): \quad    \sum_{q,q'}  \couple(i,q,j,q') \Sigmabf_{iq}(Q\cdot(\r_{\jj}, z_{\jj})_{\jj}) \left ( \Sigmabf_{jq'} (Q\cdot(\r_{\jj}, z_{\jj})_{\jj}) \right )^{\trans}
\cr \qquad \qquad=  \sum_{q,q'}\couple(i,q,j,q')Q\Sigmabf_{iq}((\r_{\jj}, z_{\jj})_{\jj}) \left [ \Sigmabf_{jq'} ((\r_{\jj}, z_{\jj})_{\jj}) \right ]^{\trans}Q^{T},
}
\end{equation}


\subsection{Permutation Equivariance}\label{app:coupling-condition}

Equivariance of $\Gammabf_{ij}$ with respect to $S_{\nat}$ in the sense of (\ref{eq:gamma:permeq}) holds, if the coupling function $\couple$ is invariant under any permutations $\sigma \in S_{\nat}$, i.e., 
\[
\forall \sigma \in S_{\nat} ~ \forall \, i,q,j,q': \quad \couple(\sigma i, \sigma q, \sigma j, \sigma q') =   \couple(i,q,j,q').
\]
This result follows because
 \begin{eqnarray*}
\Gammabf_{ij} \left ( (\r_{\sigma \jj}, z_{\sigma \jj})_{\jj} \right ) 
& =  
\sum_{q,q'} \couple(i,q,j,q')\Sigmabf_{iq}\left ( (\r_{\sigma \jj}, z_{\sigma \jj})_{\jj} \right )\left ( \Sigmabf_{iq}\left ( (\r_{\sigma \jj}, z_{\sigma \jj})_{\jj} \right ) \right )^{\trans}\\
& = 
\sum_{q,q'} \couple(i,q,j,q')\Sigmabf_{\sigma i \sigma q}\left ( (\r_{ \jj}, z_{\jj})_{\jj} \right )\left ( \Sigmabf_{\sigma i \sigma q}\left ( (\r_{ \jj}, z_{\jj})_{\jj} \right ) \right )^{\trans}\\
& = 
\sum_{q,q'} \couple(i,\sigma^{-1}q,j,\sigma^{-1}q')\Sigmabf_{\sigma i q}\left ( (\r_{ \jj}, z_{\jj})_{\jj} \right )\left ( \Sigmabf_{\sigma i q}\left ( (\r_{ \jj}, z_{\jj})_{\jj} \right ) \right )^{\trans}\\
&= \Gammabf_{\sigma i \sigma j} \left ( (\r_{\jj}, z_{\jj})_{\jj} \right ),
\end{eqnarray*}
where the last equality holds provided that 
\[
\forall \sigma \in S_{\nat} ~ \forall \, i,q,j,q': \couple(i,\sigma^{-1}q,j,\sigma^{-1}q') =  \couple(\sigma i,q, \sigma j,q'),
\]
which is equivalent to $\couple$ being invariant under any permutation $\sigma\in S_{\nat}$. 

\section{Alternative noise-couplings}\label{app:alternative-coupling}
Besides the pair-wise coupling ${\mathbb{E}}[\dot{\W}_{iq} \dot{\W}_{jq'}^{\trans}] = \delta_{i q' } \delta_{jq} {\bf I}_{m}$, discussed in the main text, there are other couplings for which (\ref{eq:gamma:permeq}) holds by virtue of the fact that the corresponding coupling function is invariant under all permutations $\sigma \in S_{\nat}$. 
Below, we provide an incomplete list of natural alternative coupling that satisfy (\ref{eq:gamma:permeq}):
\begin{enumerate}
\item The {\em row-wise coupling} of the white-noise vectors, ${\mathbb{E}}[\dot{\W}_{iq} \dot{\W}_{jq'}^{\trans}] = \delta_{q q'}  {\bf I}_{m}$, yields the following concrete form of the friction tensor
\begin{equation}\label{eq:fd:3:cc}
\Gammabf_{ij} 
= \sum_{q=1}^{\nat}\Sigmabf_{iq} \left ( \Sigmabf_{jq} \right)^{\trans}.
\end{equation}
\item The {\em column-wise coupling}  of the white-noise vectors,  $ {\mathbb{E}}[\dot{\W}_{iq} \dot{\W}_{jq'}^{\trans}] = \delta_{i j}  {\bf I}_{m}$, yields the following concrete form of the friction tensor
\begin{equation}\label{eq:fd:3:rc}
\Gammabf_{ij} 
= \delta_{ij} \sum_{q=1}^{\nat}\sum_{q'=1}^{\nat} \Sigmabf_{iq} \left ( \Sigmabf_{iq'}\right )^{\trans}.
\end{equation}
\item The {\em minimal coupling} of the white-noise vectors, ${\mathbb{E}}[\dot{\W}_{iq} \dot{\W}_{jq'}^{\trans}] = \delta_{ij} \delta_{q q'}  {\bf I}_{m}$, yields the following concrete form of the friction tensor
\begin{equation}\label{eq:fd:3:cc}
\Gammabf_{ij} 
= \delta_{ij}\sum_{q=1}^{\nat}\Sigmabf_{iq}\left(\Sigmabf_{iq}\right)^{\trans}.
\end{equation}
\item The {\em maximal coupling} of the white-noise vectors, ${\mathbb{E}}[\dot{\W}_{iq} \dot{\W}_{jq'}^{\trans}] =  {\bf I}_{m}$, yields the following concrete form of the friction tensor
\begin{equation}\label{eq:fd:3:cc}
\Gammabf_{ij} 
=\sum_{q=1}^{\nat}\sum_{q'=1}^{\nat}\Sigmabf_{iq}\left(\Sigmabf_{jq'}\right)^{\trans}.
\end{equation}
\end{enumerate}
\begin{figure}
\begin{center}
\includegraphics[width=.9\linewidth]{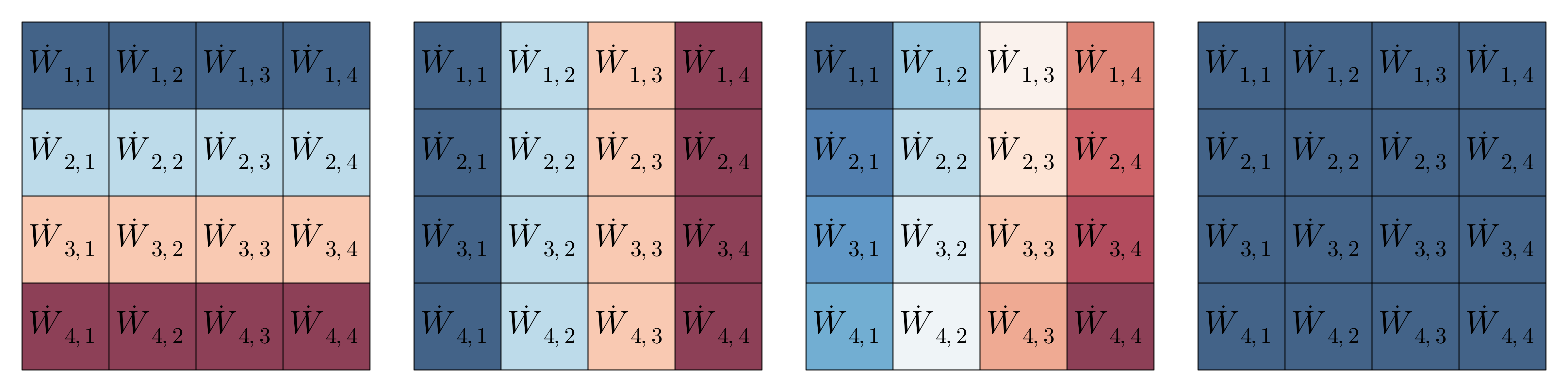}
\caption{Illustrations of alternative couplings in the case of a 4-particle system, i.e., $\nat=4$. Noise vectors with same background color are identical and thus coupled. Noise vectors with different background colors are independent. From left to right: row-wise coupling, column-wise coupling, minimal coupling and maximal coupling.} 
\end{center}
\end{figure}
\section{Atomic Cluster Expansion representation} \label{sec:pair-env-expansion}
\subsection{Equivariant ACE on pair-environments}\label{app:pair-ace}

There are various ways of constructing descriptors on pair-environments (see e.g. \cite{nigam2022equivariant,zhang2022equivariant}) that are of the functional form (\ref{eq:model-func-form}). Here, we consider an approach 
based on a smooth transformation of the local coordinates  $\r_{ij}$ and $\{ \r_{ijk}\}_{k \in \neigh(i,j)}$ of the pair environment of the atom pair $(i,j)$ into  coordinates $\{ \tr_{ijk}\}_{k \in \neigh(i,j) \cup \{j\}}$, which can be interpreted as local coordinates of a site-environment of the atom $i$. 

Specifically, in the case of the ellipsoid cutoff $E_{\lambda,\mu,r_{\rm cut}}(\r_{i},\r_{j})$, we transform coordinates and states 
as follows:
 \begin{eqnarray}
\tr_{ijk} = \cases{
 r_{\rm cut}^{-1} \left ( \r_{ijk} - (1-\mu^{-1})\mathcal{P} \r_{ijk} \right ),&  $k \in \neigh(i,j)$,\\
 e_{{\rm max}}^{-1}\r_{ij}, & $k = j$,
},
\end{eqnarray}
with $\mathcal{P}$ as defined in (\ref{eq:proj-center}),  and
 \begin{eqnarray}\label{eq:trans:z}
\tz_{k} =  \cases{
z_{k},& $k \in \neigh(i,j)$,\\
b,& $k = j$,
},
\end{eqnarray}
where $b$ is an artificial element introduced only for the purpose of a convenient implementation, distinct from the chemical element types in $\mathcal{Z}$. The significance of this coordinate transformation lies in the fact that it allows to formally rewrite $\model^{(l,p)}$ as a function $ \widetilde{\model}^{(l,p)}$ of atomic site environments, i.e., 
 \begin{eqnarray}\label{eq:pair-atom-rep}
\hspace{-2cm} \model^{(l,p)}\left (z_i, z_j,\r_{ij}, \{ (\r_{ijk}, z_{k})\}_{k \in \neigh(i,j)}\right ) 
 &= \widetilde{\model}^{(l,p)}\left( z_{i}, z_{j}, \{ ( \tr_{ijk}, \tz_{k})\}_{k \in \neigh(i,j) \cup \{j\}} \right ).
\end{eqnarray}
Assigning the distinct chemical type $b$ to the atom $j$ as in (\ref{eq:trans:z}) allows for permutation invariance for permutations that otherwise relabel the indices of the atom pair $(i,j)$ to be broken.

The starting point in the construction of an ACE of functions of the functional form of  $\widetilde{\model}^{(l,p)}$ are modified {\em one-particle basis functions} of the form
\begin{equation}
    \phi_{znlm}(z_{i},z_{j},\tz_{k}, \tilde{\bm{r}}_{ijk})=  \delta_{z\tz_{k}}R_{nl}(z_{i},z_{j},\tz_{k},\td_{ijk})Y_{l}^{m}(\hat{\tr}_{ijk}) .
\end{equation}
with $\td_{ijk} = |\tr_{ijk}|, \hat{\tr}_{ijk} = \tr_{ijk}/\td_{ijk}$, and $Y_{l}^{m}$ denotes the spherical harmonic of degree $l$ and order $m$, and the collections of functions $R_{nl}(z_{i},z_{j},\tz_{k},\cdot), n\in \mathbb{N}$, form a suitable radial basis for each $l$. The specific form of this radial basis may vary depending on the element types $z_{i},z_{j} \in \mathcal{Z}$ of the atom pair $(i,j)$ and the chemical element type $\tilde{z}_{k} \in \mathcal{Z} \cap \{b\}$ of the atom $k \in \mathcal{N}(i,j) \cap \{j\}$ for whose (transformed) displacement $\tr_{ijk}$ the one-particle basis is evaluated. 

These one-particle basis functions are then combined to form {\em pair-site basis functions}
\begin{equation}
    A_{znlm}\left (z_{i},z_{j}, \{ ( \tr_{ijk}, \tz_{k})\}_{k \in \neigh(i,j) \cup \{j\}} \right) 
    = \sum_{k \in \neigh(i,j) \cup \{j\}}  \phi_{znlm}(z_{i},z_{j},\tz_{k}, \tr_{ijk}).
\end{equation}
To obtain higher-body order permutation-invariant basis functions, products of one-particle basis functions are constructed 
\begin{equation}
    A_{\bm{\upsilon}}=\prod^{\nu}_{t=1} A_{\upsilon_{t}},
\end{equation}
where $\upsilon = znlm$, $\bm{\upsilon}=(\upsilon_{1},\dots,\upsilon_{\nu})$ and $\nu$ refers to correlation order, corresponding to a ($\nu$+1)-body basis function. In the selection of $\bm{\upsilon}$ tuples, care must be taken that the ``spurious'' element $b$ occurs at most once.
Further constraints on the selection of $\bm{\upsilon}$ may be imposed to enforce additional symmetries or specific properties of the basis functions. For example, in the electronic friction application considered in this article hydrogen molecules do not incur any friction if the two respective hydrogen atoms are not within a certain distance of the heat-bath. To enforce this property, we exclude all $\bm{\upsilon}$ of the form $\bm{\upsilon} = (bnlm)$.  

A basis $B^{i}_{\eta_{\nu}}$, enumerated by $\eta_{\nu}$, with pre-specified equivariance properties is obtained from the basis functions $A_{\bm\upsilon}$ by symmetrization, which reduces to a linear projection of the following form
\begin{equation}
    B^{(l)}_{\eta_{\nu}}=\sum_{\bm{\upsilon}} C^{(l)}_{\eta_{\nu}\bm{\upsilon}} A_{\bm{\upsilon}} ,
\end{equation}
where the coupling coefficients $C^{(l)}_{\eta_{\nu}\bm{\upsilon}}$ with  $l=0,1,2;$ are generalized Clebsch-Gordan coefficients that are suitably constructed to take values in $\RR$, $\RR^{3}$, $\RR^{3\times 3}$ to obtain scalar, vector and matrix-valued basis functions, respectively, that satisfy the correct equivariance symmetries with respect to the orthogonal group $O(3)$:
 \begin{eqnarray*}
\forall Q \in O(3): \quad  & B^{(0)}_{\eta_{\nu}}(z_{i},z_{j},\{(Q\tr_{ijk},\tz_{k})\})&= B^{(0)}_{\eta_{\nu}}(z_{i},z_{j},\{(\tr_{ijk},\tz_{j})\}),\\
\forall Q \in O(3): \quad  & B^{(1)}_{\eta_{\nu}}(z_{i},z_{j},\{(Q\tr_{ijk},\tz_{k})\})& = QB^{(1)}_{\eta_{\nu}}(z_{i},z_{j},\{(\tr_{ijk},\tz_{j})\}),\\ 
\forall Q \in O(3): \quad  & B^{(2)}_{\eta_{\nu}}(z_{i},z_{j},\{(Q\tr_{ijk},\tz_{k})\})& = QB^{(2)}_{\eta_{\nu}}(z_{i},z_{j},\{(\tr_{ijk},\tz_{j})\})Q^{T}.
 \end{eqnarray*}

\section{Representation of Vector-Equivariant Components by Matrix-Equivariant Components } \label{sec:app:mv-equ} 
The following observation justifies using heat bath models where the diffusion tensor is solely comprised of matrix-equivariant components:

Consider the diffusion tensor $\Sigmabf_{ij}$ with decomposition as specified in (\ref{eq:dt-decomp}). By replacing the components $\Sigmabf^{(1,p)}_{ij}, p=1,\dots,m_{1}$, by components $\Sigmabf^{(1,p)}_{ij} \left (\Sigmabf^{(1,p)}_{ij}\right)^{T}, p=1,\dots,m_{1}$, we obtain a new diffusion tensor 
\[
{\Sigmabf}_{ij} = [\tilde{\Sigmabf}^{(2)}_{ij}, \hat{\Sigmabf}^{(2)}_{ij}],
\]
where $\tilde{\Sigmabf}^{(2)}_{ij} =  \left [ \Sigmabf^{(1,p)}_{ij} \left(\Sigmabf^{(1,p)}_{ij} \right)^{T} \right ]_{p=1}^{m_{1}}$ and $ \hat{\Sigmabf}^{(2)}_{ij} = \Sigmabf^{(2)}_{ij}$, which, when employed with the same coupling as  $\Sigmabf_{ij}$ results in an identical friction tensor $\Gammabf_{ij}$. This observation is a direct consequence of the following lemma.

\paragraph{Lemma E.1}
Let $\mathcal{A}$ be a vector-equivariant function on atomic configurations taking values in $\RR^{3}$. Define $\mathcal{B} := \mathcal{A} \hat{\mathcal{A}}^{T}$, 
where
\[
\hat{\mathcal{A}}(\r_{i}, z_{i})_{i}) = \cases{ \mathcal{A}((\r_{i}, z_{i})_{i})/ \left \|\mathcal{A}(\r_{i}, z_{i})_{i})  \right \|_{2},&  $\mathcal{A}((\r_{i}, z_{i})_{i}) \neq (0,0,0)^{T}$, \\
(0,0,0)^{T}, & \text{otherwise}.
}
 \]
with $\left \| \; \cdot \; \right \|_{2}$ denoting the standard Euclidean norm in $\RR^{3}$.
The $\mathbb{R}^{3\times 3}$-valued function $\mathcal{B}$ is matrix-equivariant and satisfies $\mathcal{B}\mathcal{B}^{T} = \mathcal{A}\mathcal{A}^{T}$.\\
{\it Proof: } 
Since $ \hat{\mathcal{A}}$ is of unit length we have $\hat{\mathcal{A}}^{T} \hat{\mathcal{A}} = 1$, thus
\[
\mathcal{B} \mathcal{B}^{T} =  \mathcal{A} \hat{\mathcal{A}}^{T} \left ( \mathcal{A}\hat{\mathcal{A}}^{T} \right )^{T} =   \mathcal{A} \hat{\mathcal{A}}^{T}\hat{\mathcal{A}} \mathcal{A}^{T}= \mathcal{A}  \mathcal{A}^{T}.
\]

\section{Internal coordinate transformation of electronic friction tensor for diatomic molecule} \label{sec:app:eft_transform}
Internal coordinates facilitate the interpretation of the evolution of elements of the friction tensor during molecular dynamics of diatomic adsorbates. Both Cartesian $(x_1, y_1, z_1, x_2, y_2, z_2)$ and internal $(d, \theta, \phi, X, Y, Z)$ coordinates are shown schematically on the right side of Fig.~\ref{fig:validation_noau_model}. In the internal coordinate system, $d$ is the distance between adsorbate atoms of a diatomic molecule, $\theta$ and $\phi$ are the polar and azimuthal angles, and $(X, Y, Z)$ are the coordinates of the centre of mass of the adsorbate atoms. The transformation of the EFT, $\bm{\Gamma}$, from the Cartesian to internal coordinates is given by 
\begin{equation} \label{eq:int_coords_transf}
        \bm{\Gamma}_{\mathrm{int}} = \bm{U}^T \bm{\Gamma}_{\mathrm{cart}} \bm{U},
\end{equation}
where $\bm{U}$ is the transformation matrix between the coordinate systems of the diatomic molecule (here NO), 
  \begin{equation}
    \hspace{-1cm}
    \bm{U} = \left[\begin{array}{lrrrrr}
        \frac{x_N - x_O}{r} & \frac{(x_O - x_N)(z_O - z_N)}{r^2 r'} & \frac{y_N - y_O}{r'^2} & m_N/M & 0 & 0\\
        \frac{y_N - y_O}{r} & \frac{(y_O - y_N)(z_O - z_N)}{r^2 r'} & \frac{x_O - x_N}{r'^2} & 0 & m_N/M & 0\\
        \frac{z_N - z_O}{r} & \frac{-r'}{r^2} & 0 & 0 & 0 & m_N/M\\
        \frac{x_O - x_N}{r} & \frac{(x_N - x_O)(z_O - z_N)}{r^2 r'} & \frac{y_O - y_N}{r'^2} & m_O/M & 0 & 0\\
        \frac{y_O - y_N}{r} & \frac{(y_N - y_O)(z_O - z_N)}{r^2 r'} & \frac{x_N - x_O}{r'^2} & 0 & m_O/M & 0\\
        \frac{z_O - z_N}{r} & \frac{r'}{r^2} & 0 & 0 & 0 & m_O/M\\
        
        \end{array}\right], 
  \end{equation}
where $x_a$, $y_a$, $z_a$ are Cartesian coordinates of atom \textit{a},\\
$r = \sqrt{(x_O - x_N)^2 + (y_O - y_N)^2 + (z_O - z_N)^2}$, $r' = \sqrt{(x_O - x_N)^2 + (y_O - y_N)^2}$, and $M = m_N + m_O$. The column vectors of the transformation matrix are normalised to conserve the trace of the friction tensor during projection. The resulting EFT contains the diagonal values, that are associated with the electronic friction in the internal coordinates $(d, \theta, \phi, X, Y, Z)$, and non-diagonal values correspond to the coupling between the two internal coordinates.

\section*{Acknowledgements}

WS  was supported by a Leverhulme Trust Research Project Grant (RPG-2019-078). RJM acknowledges support through the UKRI Future Leaders Fellowship programme (MR/S016023/1 and MR/X023109/1), and a UKRI frontier research grant (EP/X014088/1). CO was supported by NSERC Discovery Grant GR019381 and NFRF Exploration Grant GR022937.

High-performance computing resources were provided via the Scientific Computing Research Technology Platform of the University of Warwick and the EPSRC-funded HPC Midlands+ computing centre for access to Sulis (EP/P020232/1).

\section*{References}
\bibliographystyle{iopart-num}
\bibliography{friction-tensor-refs.bib}

\end{document}